\documentclass[floats,floatfix,showpacs,amssymb,prd,twocolumn,superscriptaddress,nofootinbib,nolongbibliography,reprint]{revtex4-1}

\usepackage{amssymb,amsmath,verbatim,mathtools,needspace,enumitem,etoolbox,graphicx,physics,microtype,afterpage,xspace,tabularx,lmodern,multirow}
\usepackage{gensymb}
\usepackage[normalem]{ulem}
\usepackage[dvipsnames, usenames]{xcolor}
\usepackage{xr-hyper}
\definecolor{linkcolor}{rgb}{0.0,0.3,0.5}
\usepackage[unicode, colorlinks=true, linkcolor=linkcolor, citecolor=linkcolor, filecolor=linkcolor, urlcolor=linkcolor, linktocpage, breaklinks]{hyperref}
\usepackage[all]{hypcap}
\usepackage[T1]{fontenc}
\usepackage[utf8]{inputenc}
\usepackage[usenames,dvipsnames]{xcolor}
\hypersetup{colorlinks=true,citecolor=romared,linkcolor=romared,urlcolor=romared}

\definecolor{romared}{RGB}{142,0,28}

\newcommand{\be}{\begin{equation}}
\newcommand{\ee}{\end{equation}}

\def\be{\begin{equation}}
\def\ee{\end{equation}}
\newcommand{\beq}{\begin{eqnarray}}
\newcommand{\eeq}{\end{eqnarray}}

\usepackage{makecell}
\usepackage{soul}

\newcolumntype{Y}{>{\centering\arraybackslash}X}

\newcommand\prlsec[1]{\vspace{2mm}\noindent {\bf \emph{#1}}}

\begin{document}

\title{
Quadrupole instability of static scalarized black holes
}

\author{{Burkhard Kleihaus}}
\affiliation{Institut f\"ur Physik, Universit\"at Oldenburg, Postfach 2503, D-26111 Oldenburg, Germany}
\author{{Jutta Kunz}}
\affiliation{Institut f\"ur Physik, Universit\"at Oldenburg, Postfach 2503, D-26111 Oldenburg, Germany}
\author{{Tim Uterm\"ohlen}}
\affiliation{Institut f\"ur Physik, Universit\"at Oldenburg, Postfach 2503, D-26111 Oldenburg, Germany}
\author{{Emanuele Berti}}
\affiliation{Department of Physics and Astronomy, Johns Hopkins University, Baltimore, MD 21218 USA}

\date{\today}


\begin{abstract} 
The addition of a Ricci coupling to Einstein-scalar-Gauss-Bonnet theories makes general relativity a cosmological attractor.
Previous work considered a quadratic coupling function with two independent coupling constants in such theories and showed that static, spherically symmetric, spontaneously scalarized black holes are radially stable beyond a critical value of the Ricci coupling constant.
Here we demonstrate that these black holes are affected by a quadrupole instability which leads to two new branches of static, axially symmetric scalarized black holes.
We discuss the properties of these solutions and provide embedding diagrams.
\end{abstract}

\maketitle

\prlsec{Introduction.}
In general relativity (GR), the no-hair theorems highly restrict the allowed black hole (BH) solutions and their properties~\cite{Chrusciel:2012jk,Herdeiro:2015waa}.
Real scalar fields, for instance, cannot lead to scalar hair.
The situation is different in generalized theories of gravity intended to amend various shortcomings of GR (see, e.g.,~\cite{Berti:2015itd,Saridakis:2021vue}).

A particularly attractive and well-studied class of theories contains a Gauss-Bonnet (GB) quadratic term coupled to a real scalar field.
Such Einstein-scalar-Gauss-Bonnet (EsGB) theories lead to second order equations of motion and do not feature ghosts.
In the low-energy limit of string theory, the scalar field corresponds to a dilaton~\cite{Gross:1986mw,Metsaev:1987zx}. In these Einstein-dilaton-Gauss-Bonnet theories BHs always carry scalar hair~\cite{Kanti:1995vq,Kleihaus:2011tg}, and the vacuum BHs of GR are no longer solutions. 

However, EsGB theories allowing for different (non-dilatonic) coupling functions $f(\phi)$ of the scalar field to the GB term admit GR BHs as special solutions of the field equations.
For example, theories with coupling functions quadratic in $\phi$ 
feature BH solutions with scalar hair that reduce to GR solutions for small couplings. These ``spontaneously scalarized'' BH solutions arise from a tachyonic instability of the GR BHs, due either to strong curvatures or to large spins~\cite{Doneva:2017bvd,Silva:2017uqg,Antoniou:2017acq,Antoniou:2017hxj,Cunha:2019dwb,Collodel:2019kkx,Dima:2020yac,Hod:2020jjy,Herdeiro:2020wei,Berti:2020kgk}.

Static, spherically symmetric, spontaneously scalarized BHs ``branch off'' the Schwarzschild BH solutions below a critical value of the mass (for fixed coupling constant), where the tachyonic instability produces a zero mode of the Schwarzschild BHs.
For masses lower than this critical value, the Schwarzschild BHs possess an unstable radial mode~\cite{Blazquez-Salcedo:2018jnn,Silva:2018qhn,Macedo:2019sem} and a new branch of scalarized BHs (the ``fundamental branch'') emerges. The stability of scalarized BHs under radial perturbations depends on the specific choice of the coupling function and of the scalar field potential~\cite{Blazquez-Salcedo:2018jnn,Silva:2018qhn,Macedo:2019sem}.

As shown in Refs.~\cite{Blazquez-Salcedo:2018jnn,Blazquez-Salcedo:2020rhf,Blazquez-Salcedo:2020caw},  linear mode stability applies to (most of) the fundamental scalarized branch of the extended scalar-tensor GB theories considered in Ref.~\cite{Doneva:2017bvd}.
In contrast, for a purely quadratic coupling function, $f(\phi) = \phi^2/2$, the fundamental scalarized branch is radially unstable everywhere~\cite{Blazquez-Salcedo:2018jnn}.
The fundamental branch can be made partially (radially) stable by including higher order terms or a potential. Then the radial instability sets in at a minimum of the scalarized BH mass~\cite{Silva:2018qhn,Macedo:2019sem}.

Recently, a partially radially stable fundamental branch of scalarized BHs was found in another interesting scenario~\cite{Antoniou:2020nax,Antoniou:2021zoy,Antoniou:2022agj}:
Einstein-scalar-Gauss-Bonnet-Ricci (EsGBR) theories, which include also a term coupling the scalar field with the Ricci scalar via a quadratic coupling function (but with a different coupling constant).
EsGBR theories are well motivated from a cosmological point of view, since they allow GR to be a cosmological attractor~\cite{Antoniou:2020nax}:
no fine-tuning of the scalar field in the early Universe is needed in order to have a vanishing scalar field at late times.

Here we revisit scalarized BH solutions for the EsGBR action~\cite{Antoniou:2020nax,Antoniou:2021zoy,Antoniou:2022agj}
\begin{eqnarray}  
\label{act}
 {\cal S} = \frac{1}{16 \pi} & \int & \mathrm{d}^4x \sqrt{-g}  \left[ R - \frac{1}{2}(\partial_\mu \phi)^2 \right. \nonumber\\
&& \left.  -   \frac{\phi^2}{2} \left(\frac{\beta}{2}R -\alpha R_{\mathrm{GB}}^2\right)  \right]
\end{eqnarray} 
with a real scalar field $\phi$, 
coupling constants $\alpha$ and $\beta$, and GB invariant
 $R^2_{\rm GB} = R_{\mu\nu\rho\sigma} R^{\mu\nu\rho\sigma}
- 4 R_{\mu\nu} R^{\mu\nu} + R^2$ \,.

We show that scalarized BHs in EsGBR theories have an intriguing new feature:
radially stable, static, spherically symmetric BHs on the fundamental scalarized branch develop a quadrupole instability below a critical value of the BH mass.
At this critical value of the mass the spherically symmetric solutions possess a zero mode from where two branches of static, but only {\em axially} symmetric, BH solutions arise.
Therefore, stability under radial perturbations may not necessarily imply linear mode stability.
We present the domain of existence of the new branches of BH solutions and study their physical properties.

\prlsec{General framework.}
\label{sec:framework}
Variation of the action \eqref{act} with respect to the metric yields the generalized Einstein equations
\begin{align}
    E_{\mu\nu} = G_{\mu\nu} - \frac{1}{2}T_{\mu\nu}^{\mathrm{(eff)}} = 0 \,,
\end{align}
where the effective stress-energy tensor
\begin{align}
    T_{\mu\nu}^{\mathrm{(eff)}} = T_{\mu\nu}^{(\phi)} - 2\alpha T_{\mu\nu}^{(\mathrm{GB})} + \beta T_{\mu\nu}^{(\mathrm{R})}
\end{align}
has contributions from the scalar, GB and Ricci terms, respectively.
Variation with respect to the scalar field yields the generalized Klein-Gordon equation
\begin{eqnarray}
\label{dil-eq}
\nabla^2 \phi - 
\left(\frac{\beta}{2}R -\alpha R_{\mathrm{GB}}^2\right) \phi
 =0 \,.
\end{eqnarray} 
The latter features the effective mass 
\begin{equation}
    m^2_{\rm eff} =  \frac{\beta}{2} R - \alpha R^2_{\rm GB} \,,
\end{equation}
which allows for spontaneous scalarization and for the attractive cosmological features of the model.

We find solutions corresponding to static, axially symmetric spacetimes by imposing the ansatz~\cite{Kleihaus:1997ic,Kleihaus:2000kg}
\begin{eqnarray}
\label{metric}
ds^2&=&- b e^{F_0} dt^2 + e^{F_1} \left( d r^2+ r^2d\theta^2 \right) \nonumber\\
           &+&  e^{F_2} r^2\sin^2\theta d\varphi^2 \,,
\end{eqnarray}
with ``quasi-isotropic" radial coordinate $r$, auxiliary function $ b=\left( 1- \frac{r_{\rm H}}{r} \right)^2 $ 
(here $r_{\rm H}$ is the isotropic horizon radius), and three unknown metric functions $F_0$, $F_1$, $F_2$.
These metric functions $F_i$ $(i=0,\,1,\,2$) and the scalar field $\phi$ depend only on the coordinates $r$ and $\theta$.

Based on considerations of symmetry, regularity and asymptotic flatness of the solutions, we impose the following set of boundary conditions at spatial infinity, at the horizon and on the symmetry axis, respectively:
$F_i(\infty)=0 \quad (i=0,\,1,\,2),\,\quad \phi(\infty)=0$; 
$\partial_r F_0(r_{\rm H})=1/r_{\rm H}$,
$\partial_r F_1(r_{\rm H})=-2/r_{\rm H}$,
$\partial_r F_2(r_{\rm H})=-2/r_{\rm H}$,
$\partial_r \phi(r_{\rm H})=0$; and
$\partial_\theta F_i|_{\theta={0,\pi}} =0$ $(i=0,\,1,\,2)$,
$\partial_\theta \phi|_{\theta={0,\pi}} =0$.

Assuming the scalar field to be even under parity, a series expansion at infinity leads to $\phi=Q/r+\dots$, where $Q$ represents the scalar charge.
The BH mass $M$ can be found from the metric function $g_{tt}  =-1+2M/r+\dots$.
Since all functions are even, we can limit calculations to one quadrant only and impose the boundary conditions
$\partial_\theta F_i|_{\theta=\pi/2} = 0$ $(i=0,\,1,\,2)$, $\partial_\theta \phi|_{\theta=\pi/2} = 0$ on the equatorial plane.

The horizon metric determines the Hawking temperature~\cite{Wald:1984rg}
\begin{eqnarray}
\label{TH}
T_{\rm H}=\frac{1}{2 \pi r_{\rm H}}e^{(F_0-F_1)/2}
\end{eqnarray}
and the horizon area
\begin{eqnarray}
\label{AH}
A_{\rm H}=2\pi r_{\rm H}^2\int_0^\pi d\theta 
\sin\theta e^{(F_1+F_2)/2}\,.
\end{eqnarray}
The entropy, however, differs from the GR result, where it is simply a quarter of the horizon area~\cite{Wald:1984rg}.
For EsGBR BHs there are additional contributions~\cite{Wald:1993nt}, and 
the entropy is found as the following integral over the spatial cross section of the horizon:
\begin{eqnarray}
\label{S-Noether} 
S=\frac{1}{4}\int_{\Sigma_{\rm H}} d^{2}x \sqrt{h} 
\left[ 1 - \frac{\phi^2}{2} \left(\frac{\beta}{2} -2 \alpha \tilde R \right) \right]\,,
\end{eqnarray} 
where $h$ is the determinant of the induced metric on the horizon and $\tilde R$ is the corresponding scalar curvature.

\prlsec{Numerical Results.}
\label{sec:numerics}
The numerical approach we employed previously for EsGB BHs~\cite{Collodel:2019kkx,Berti:2020kgk} is based on the finite difference solver FIDISOL/CADSOL~\cite{Schoenauer:1989,Schoenauer:1989b}.
For EsGBR BHs, we supplemented this solver by a spectral solver.
In both cases the unknown functions $(F_0, F_1, F_2, \phi)$ are obtained for given values of the coupling constants $\alpha$ and $\beta$ and of the horizon radius $r_{\rm H}$ by solving the chosen set of partial differential equations (PDEs) subject to the prescribed boundary conditions.
In particular, we employ the linear combinations of the Einstein equations $E_t^t,E_\varphi^\varphi,E_\varphi^t$ and $E_r^r+E_\theta^\theta$ for the PDEs yielding the metric functions, and we treat the equations involving $E_r^r-E_\theta^\theta$ and $E_r^\theta$ as constraints.
%
Introducing a compactified radial variable $x=1-r_{\rm H}/r$, we map the interval $[r_{\rm H},\infty)$ to the finite interval $[0,1]$.
We then discretize the equations on a nonequidistant grid in the variables $x$ and $\theta$ whose range is $0\leq x \leq 1$ and $0\leq \theta \leq \pi/2$.
The resulting estimated numerical error is of order $10^{-3}$ or less.

\begin{figure}[t]
\includegraphics[width=0.5\textwidth]{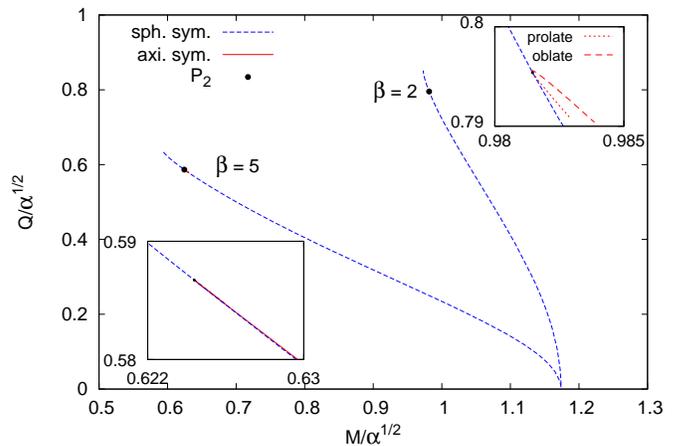}
    \caption{
Branches of scalarized BHs for $\beta=2$ and 5: scaled scalar charge $Q/\sqrt{\alpha}$ vs scaled mass $M/\sqrt{\alpha}$ for the fundamental spherically symmetric fundamental branches (dotted blue) and the axially symmetric branches 
(red) in the insets.
The critical bifurcation points $P_2$ are shown by black dots.
}
        \label{Fig:domain1}
\end{figure}

We begin our discussion by recalling the fundamental scalarized static spherically symmetric BH branches~\cite{Antoniou:2021zoy,Antoniou:2022agj}.
When $\beta$ exceeds a critical value of about 1.15, radially stable BHs arise from the bifurcation point with the Schwarzschild BHs (see e.g. Fig.~2 of Ref.~\cite{Antoniou:2022agj}).
Analogous to EsGB BHs~\cite{Silva:2018qhn,Macedo:2019sem}, for smaller values of $\beta$ these branches feature a minimum of the mass, where the radial instability sets in.
For larger values of $\beta$ the branches are radially stable, and they terminate with a solution such that a certain radicand in the horizon expansion of the scalar field vanishes~\cite{Kanti:1995vq}.
In Fig.~\ref{Fig:domain1} we show the scaled scalar charge $Q/\sqrt{\alpha}$ versus the scaled mass $M/\sqrt{\alpha}$ for two values of the coupling, $\beta =2$ and $\beta=5$.
These results agree with those of Ref.~\cite{Antoniou:2022agj}.

However, unlike previously assumed, radial stability along these branches does not imply general mode stability.
At the locations marked by black dots in Fig.~\ref{Fig:domain1} the scalarized BHs gain a zero mode with respect to quadrupole deformations, and we find two branches of axially symmetric BH solutions (red dashed and red dotted lines in the insets).

\begin{figure}[t]
\includegraphics[width=0.5\textwidth]{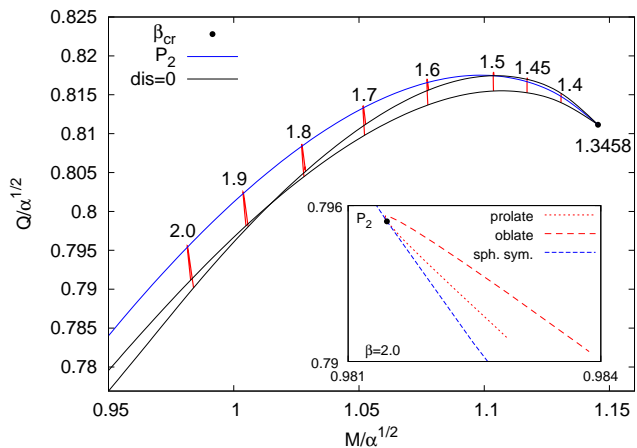}
        \caption{
Domain of existence of axially symmetric scalarized BHs: scaled scalar charge $Q/\sqrt{\alpha}$ vs scaled mass $M/\sqrt{\alpha}$ for the endpoints of the two branches 
(black,
determined by the vanishing of the respective discriminant) together with the critical points of the fundamental spherically symmetric branch 
(blue, determined by the zero mode).
The black star indicates the onset of the quadrupole instability at $\beta=1.3458$.
The numbers correspond to selected values of $\beta$.
The inset shows a zoom of the spherically and axially symmetric branches for $\beta=2$. 
}
        \label{Fig:domain2}
\end{figure}



Let us denote the bifurcation points of the axial branches by $P_2$. The blue line in Fig.~\ref{Fig:domain2} shows these bifurcation points, starting from the critical value $\beta_{\rm cr} = 1.3458$ where they first appear.  We note that for $\beta_{\rm cr} \leq \beta \lesssim 1.52$ the onset of instability occurs on the radially unstable part of the corresponding spherically symmetric branch, whereas for $\beta \gtrsim 1.52$ it occurs on the radially stable part.
We also show the endpoints of the two axial branches. These are denoted by ``$\mathrm{dis}=0$,'' since in the horizon expansion for the solutions a discriminant vanishes, and the existence of real BH solutions requires a positive sign~\cite{Berti:2020kgk}. Therefore the region between the blue line and the two black lines represents the domain of existence of the two axial branches for $M/\sqrt{\alpha} > 0.95$.

The inset in Fig.~\ref{Fig:domain2} is a zoom-in of the axial branches for $\beta=2$.
Starting at the bifurcation with the spherical branch, the upper branch (oblate, red dashed line) curiously features a small portion where the BH mass decreases, but then it bends towards larger BH masses. On the contrary, the lower branch (prolate, red dotted line) immediately moves in the direction of larger BH masses.
This curious feature is lost for larger values of $\beta$. 

\begin{figure}[t]
{\vspace*{-0.5cm}\hspace*{-4.5cm}\includegraphics[width=0.97\textwidth]{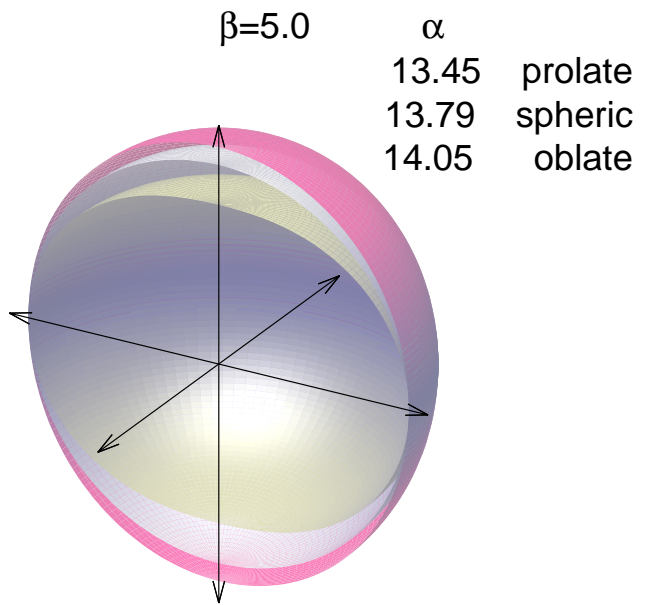}
\vspace*{-2.5cm}}
        \caption{
Embedding of the horizon of scalarized BHs for $\beta=5$: critical spherical BH ($\alpha = 13.79$), 
endpoints of upper ($\alpha = 14.05$) and lower ($\alpha = 13.45$) axial branch for
fixed circumferential radius of the horizon.
}
        \label{Fig:domain}
\end{figure}

At the endpoints of the axial branches, their deformation becomes maximal.
The deformation of the two axial branches is illustrated in Fig.~\ref{Fig:domain},
where we show embeddings of the BH horizons at the endpoints of the two axial branches for $\beta=5$ and compare them with the spherical horizon.
Clearly, along the lower (dotted) axial branch the deformation becomes prolate, while along the upper (dashed) axial branch it becomes oblate.

\begin{figure*}[t]
\includegraphics[width=\columnwidth]{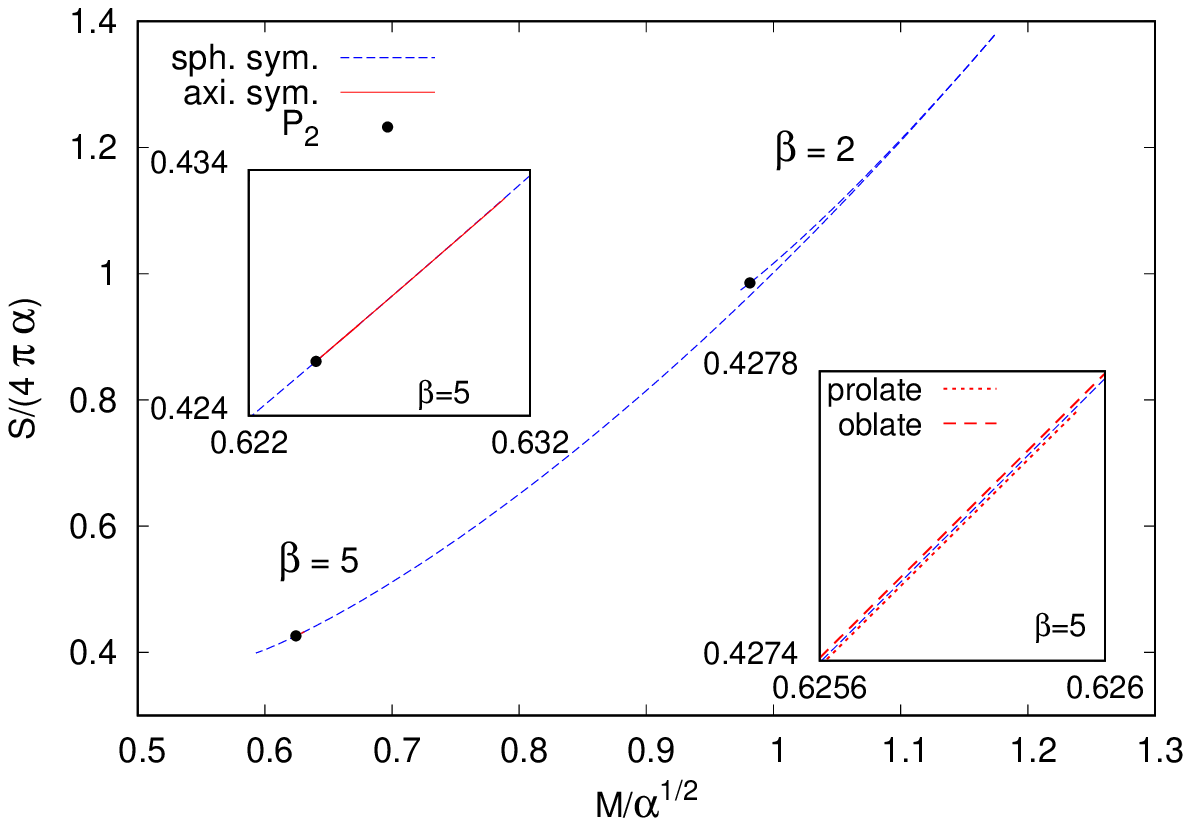}
\includegraphics[width=\columnwidth]{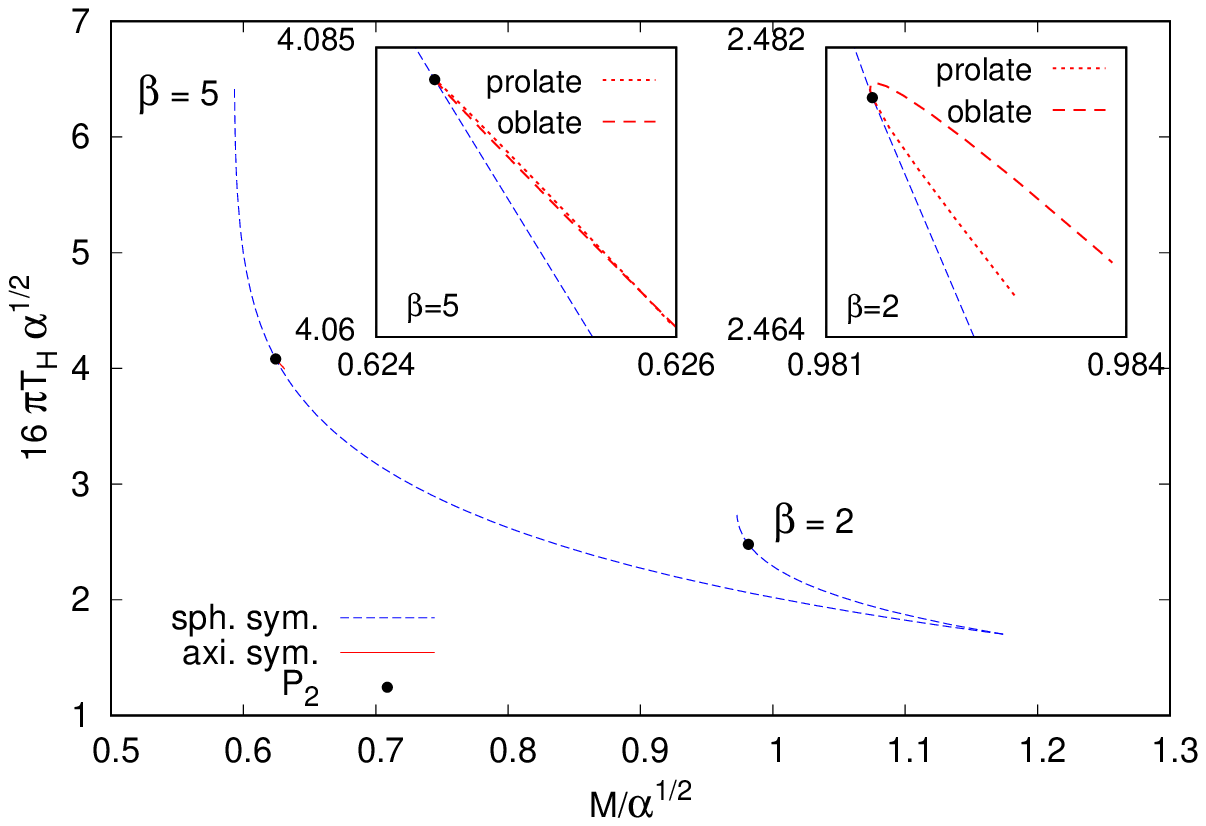}
    \caption{
Scaled entropy $S/(4\pi \alpha)$ (left panel) and scaled Hawking temperature $16\pi T_H\sqrt{\alpha}$ (right panel) versus the scaled mass $M/\sqrt{\alpha}$ for the spherically symmetric fundamental branches (dotted blue) of scalarized BHs with $\beta=2$ and $\beta=5$. 
Black dots mark the critical bifurcation points.
The prolate and oblate axially symmetric branches (red dotted and red dashed lines, respectively) are shown in the insets.
}
        \label{Fig:domain3}
\end{figure*}

In Fig.~\ref{Fig:domain3} we turn to thermodynamic properties, plotting the scaled entropy $S/(4\pi \alpha)$ (left panel) and the scaled Hawking temperature $16\pi T_H\sqrt{\alpha}$ (right panel) versus the scaled mass $M/\sqrt{\alpha}$ for the spherical branches with $\beta=2$ and $\beta=5$. 

Let us focus first on the left panel. The entropy of the axial branches for $\beta=5$ (shown in the insets) is very close to the entropy along the spherical solutions, but a high magnification (bottom right inset) uncovers small deviations: the prolate lower branch (dotted) has slightly higher entropy than the spherical branch, while the oblate upper branch (dashed) has slightly smaller entropy. Therefore the BHs on the prolate branch are, if only slightly, entropically favored.
The closeness of the entropy of the spherical and axial branches is somewhat surprising, because the scaled horizon area $A_{\rm H}/\alpha$ of the spherical branch is considerably larger than the area of the axial branches away from the bifurcation for BHs with the same scaled mass $M/\sqrt{\alpha}$.

The right panel of Fig.~\ref{Fig:domain3} shows that the temperature $T_{\rm H}\sqrt{\alpha}$ of the scalarized BHs is a decreasing function of $M/\sqrt{\alpha}$, as in the case of Schwarzschild BHs.
Both axial branches possess larger $T_{\rm H}\sqrt{\alpha}$ than the spherical branch.
The temperature of the axially symmetric BHs is larger than the temperature of spherically symmetric BHs on the radially stable branch, while it is smaller than the temperature of spherically symmetric BHs on the radially unstable branch (not shown in the figure).
For large enough values of $\beta$ (including the value 2, shown in the top right inset), there is still a part of the branch of oblate BHs (dashed red line) where the temperature increases with mass.
This feature seems to disappear for $\beta \gtrsim 3$.

\prlsec{Conclusions.}
\label{sec:conclusions}
We have investigated spontaneously scalarized, static BH solutions in EsGBR theories with quadratic coupling functions. 
These cosmologically motivated theories allow for branches of static spherically symmetric BHs that are (in part) stable with respect to radial perturbations~\cite{Antoniou:2020nax,Antoniou:2021zoy,Antoniou:2022agj}.
We have found that radial stability does not, in general, imply perturbative mode stability.
In particular, we have demonstrated the occurrence of a quadrupole instability along the spherical branches for a wide range of coupling constants.

Two distinct, axially deformed branches of BHs arise starting from a zero mode of the spherical BHs with respect to quadrupole deformations.
The lower branch is prolate and entropically favored, while the upper branch is oblate and entropically disfavored relative to the spherical branch, although the difference in entropy is very small (the differences in horizon area and temperature are larger).
These static axial branches of BHs represent new counterexamples to Israel's theorem~\cite{Israel:1967za}.

The occurrence of a quadrupole instability might be attributed to the presence of the Ricci coupling.
However, we cannot exclude the possibility that such an instability may occur in simpler EsGB theories with higher order terms in the coupling function or with a nonzero potential~\cite{Silva:2018qhn,Macedo:2019sem}. This is an interesting topic for further study.

We have considered the possibility of a dipole instability of EsGBR BHs, because Schwarzschild BH instabilities in EsGB theories involve first for the monopole (i.e., the radial instability) and then the dipole, before the quadrupole instability occurs~\cite{Collodel:2019kkx}. However, our numerical investigations did not reveal any dipole instability.

A perturbative study of the mode stability of the new axial branches will be technically challenging, and their nonlinear evolution is particularly interesting. 
Before tackling these difficult questions, it should be possible to understand various other physical properties of these solutions, such as their shadow. 
The axial symmetry of the static spacetimes should lead to deformations of the shadow, but without distinction with respect to co- and counter-rotation. A comparison with observations could set constraints on the coupling constants of the theory~\cite{Akiyama:2019eap}.

The static axially symmetric branches can be set into rotation, just like the spherically symmetric branches.
This implies interesting overlapping domains of existence for rotating BHs in EsGBR theories, that will be presented elsewhere~\cite{Berti:2023}.

\prlsec{Acknowledgments.}
 B.K. and J.K. gratefully acknowledge support by the
DFG Research Training Group 1620  \textit{Models of Gravity}
and DFG project Ku612/18-1.
E.B. is supported by NSF Grants No. AST-2006538, PHY-2207502, PHY-090003 and PHY-20043, and NASA Grants No. 19-ATP19-0051, 20-LPS20-0011 and 21-ATP21-0010.
This research project was conducted using computational resources at the Maryland Advanced Research Computing Center (MARCC). 
This work has received funding from the European Union’s Horizon 2020 research and innovation programme under the Marie Skłodowska-Curie grant agreement No. 690904. This research project was conducted using computational resources at the Maryland Advanced Research Computing Center (MARCC).

\bibliography{references}

\end{document}